\documentclass{aa}
\usepackage{longtable}
\usepackage{graphicx,natbib,txfonts,color}
\bibpunct{(}{)}{;}{a}{}{,}
\newcommand{\ang}{{\AA}}

\newcommand{\kms}{km\,s$^{-1}$}

\newcommand{\nulacht}{\object{IRAS 08544-4431}\,}
\newcommand{\hd}{\object{IRAS 12222-4652}\,}
\newcommand{\entra}{\object{EN TrA}\,}
\newcommand{\eenvijf}{\object{IRAS 15469-5311}\,}
\newcommand{\eennegen}{\object{IRAS 19125+0343}\,}
\newcommand{\eennegenb}{\object{IRAS 19157-0247}\,}

\begin{document}


\title{Post-AGB stars with hot circumstellar dust: binarity of the 
low-amplitude pulsators.
\thanks{based on observations collected with the Flemish 1.2m Mercator 
telescope at Roque de los Muchachos (Spain), the Swiss 1.2m Euler telescope 
at La Silla (Chile) and the 0.5m and 0.75m telescopes at SAAO
(South-Africa).} }

\author{Hans Van Winckel\inst{1}
\and Tom Lloyd Evans\inst{2}
\and Maryline Briquet\inst{1}\thanks{Postdoctoral fellow of the Fund for
        Scientific Research, Flanders}
\and Peter De Cat\inst{1,3}
\and Pieter Degroote\inst{1}
\and Wim De Meester\inst{1}
\and Joris De Ridder\inst{1}$^{**}$
\and Pieter Deroo\inst{1,4}
\and Maarten Desmet\inst{1}
\and Rachel Drummond\inst{1,5}
\and Laurent Eyer\inst{1,6}
\and Martin A.T. Groenewegen\inst{1,3}
\and Katrien Kolenberg\inst{1,7}
\and David Kilkenny\inst{9}
\and Djazia Ladjal\inst{1}
\and Karolien Lefever\inst{1,5}
\and Thomas Maas\inst{1}
\and Fred Marang\inst{10}
\and Peter Martinez\inst{10}
\and Roy H. {\O}stensen \inst{1}
\and Gert Raskin\inst{1}
\and Maarten Reyniers\inst{1,8}
\and Pierre Royer\inst{1}
\and Sophie Saesen\inst{1}\thanks{PhD student of the Fund for
        Scientific Research, Flanders}
\and Katrien Uytterhoeven\inst{1,11}
\and Jan Vanautgaerden\inst{1}
\and Bart Vandenbussche\inst{1}
\and Francois van Wyk\inst{10}
\and Maja Vu{\v c}kovi{\'c}\inst{1}
\and Christoffel Waelkens\inst{1}
\and Wolfgang Zima\inst{1}}

\offprints{H. Van Winckel, Hans.VanWinckel@ster.kuleuven.be}

\institute{ Instituut voor Sterrenkunde, K.U.Leuven, Celestijnenlaan 200B,
B-3001 Leuven, Belgium 
\and SUPA, School of Physics and Astronomy, University of St Andrews, North Haugh,
St Andrews, Fife, Scotland KY16 9SS 
\and  Royal Observatory of Belgium, Ringlaan 3, 1180 Brussel, Belgium
\and Jet Propulsion Laboratory, Caltech, 4800 Oak Grove Drive, Pasadena, 
CA 91109, USA
\and Belgian Institute for Space aeronomy, Ringlaan 3, 1180 Brussels, Belgium
\and Observatoire de Gen\`eve, CH 1290 Sauverny, Switzerland
\and Institut f\"ur Astronomie der Universit\"at Wien, T\"urkenschanzstrasse, 
17, A - 1180 Vienna, Austria
\and The Royal Meteorological Institute of Belgium, Department Observations,
Ringlaan 3, 1180 Brussels, Belgium
\and Dept. of Physics, University of the Western Cape, Private Bag X17, 
Bellville 7535, Western Cape, South Africa
\and South African Astronomical Observatory, P.O.Box 9, Observatory
7935, South Africa 
\and Laboratoire AIM, CEA/DSM-CNRS-Universit\'e Paris Diderot; CEA,
IRFU, SAp, centre de Saclay, F-91191, Gif-sur-Yvette, France
}

\date{Received  / Accepted}

\authorrunning{H. Van Winckel et al.}
\titlerunning{Binarity of post-AGB stars}

\abstract
{The influence of binarity on the late stages of
  stellar evolution.}
{While the first binary post-AGB stars were serendipitously discovered,
the distinct characteristics of their Spectral Energy Distribution
(SED) allowed us to launch a more
systematic search for binaries.  We selected post-AGB objects which show 
a broad dust excess often starting already at H or K, pointing to the presence 
of a gravitationally bound dusty disc in the system.
We started a very extensive multi-wavelength study of those systems
and here we report on our radial velocity and photometric monitoring results 
for six stars of early F type, which are pulsators of small amplitude.
}
{To determine the radial velocity of low signal-to-noise time-series, 
 we constructed dedicated auto-correlation masks based on
  high signal-to-noise spectra, used in our published chemical
  studies. The radial velocity variations were subjected to detailed
  analysis to differentiate between pulsational variability and
  variability due to orbital motion. When available, the photometric
  monitoring data were used to complement the time series of radial
  velocity data and to establish the nature of the pulsation.
  Finally orbital minimalisation was performed to constrain the
  orbital elements.}
{All of the six objects are binaries, with orbital periods
  ranging from 120 to 1800 days. Five systems have non-circular
  orbits. The mass functions range from 0.004 to 0.57 M$_{\odot}$ and
  the companions are likely unevolved objects of (very) low initial mass.
 We argue that these binaries must have been subject to
  severe binary interaction when the primary was a cool supergiant.
Although the origin of the circumstellar disc is not well understood, 
the disc is generally believed to be formed during this strong
interaction phase. The eccentric orbits of these highly evolved
objects remain poorly understood. In one object the line-of-sight is
grazing the edge of the puffed-up inner rim of the disc.}
{These results corroborate our earlier statement that evolved objects in binary
stars create a Keplerian dusty circumbinary disc. With the measured
orbits and mass functions we conclude that the circumbinary discs seem to
have a major impact on the evolution of a significant fraction of binary systems.} 


\maketitle


\section{Introduction}\label{sect:intro}

Post-AGB stars are low- and intermediate-mass stars that evolve
rapidly from the Asymptotic Giant Branch (AGB) toward the Planetary
Nebulae phase (PNe), before cooling down as a white dwarf. The
processes which govern the transition between the symmetry in the AGB
outflows and the observed large variety in geometries of proto-PNe and
PNe, are still badly understood \citep[][and references therein]{balick02, 
sahai07}. During the transition time, the star and its circumstellar 
envelope must be subject to fundamental and rapid changes in structure, 
mass-loss mode and geometry. 
It is now generally acknowledged that binary interaction in
these evolved stars could be responsible for a large fraction of the
observed asymmetric nebular morphologies, but the direct observational 
evidence for binarity in PNe is weak \citep{moe06, zijlstra07}. 
A recent analysis of the lightcurves obtained by the OGLE microlensing 
survey towards the Galactic Bulge, revealed a close binary fraction of 12-21 \%
\citep{miszalski09}. These data are only sensitive to the detection of
systems which were subject to a strong orbital decrease during interaction.

Testing of the (wide) binary hypothesis is severely hampered by the lack of
observational information on binarity in PNe and the often very obscured
proto planetary nebulae (PPNe), and also by our poor theoretical
understanding of AGB evolution in binary systems.  A major shortcoming
is that detailed studies of individual, often spectacular, evolved objects
prevail and systematic studies of a homogeneous sample are lacking.
To study late stellar evolution in binary systems, optically bright,
less obscured post-AGB stars are ideal candidates and in recent years
it has become clear that binarity is indeed very common \citep{vanwinckel07}.

The first binary post-AGB stars were serendipitously discovered and
turned out to have distinct observational characteristics, which include
broad dust excesses often starting already at H or K, pointing to the
presence of both hot and cool dust around the system.  It was
postulated that this indicates the presence of gravitationally-bound
circumstellar material in the system \citep[][and references
therein]{vanwinckel03}.  The most famous example is the Red Rectangle,
for which the Keplerian kinematics have been resolved by
interferometric CO measurements \citep{bujarrabal05}.

These characteristics of their SEDs allowed us to launch a more
systematic search for candidate binaries. In \citet{deruyter06} we
presented 51 objects and the number of known examples has increased since. 
The total sample is now a fair proportion of the 326 ``very likely
post-AGB stars'' known in the Galaxy \citep{szczerba07}. Our selection
criteria were focused on the infrared colours and resulted in
including the RV\,Tauri stars with a dust excess detected by IRAS, the
known binary post-AGB stars and the newly characterised objects which
were selected by \citet{lloydevans99} for their position in the
'RV\,Tauri' box in the IRAS $[12]-[25]$, $[25]-[60]$ two-color diagram
\citep[see also][]{maas03}.

\begin{figure}
\vspace{0cm}
\hspace{0cm}
\resizebox{\hsize}{!}{\includegraphics{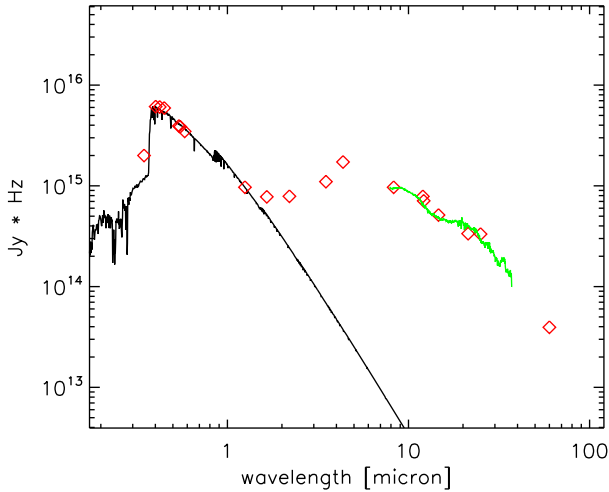}}
\caption{\label{fig:sed} The SED of \eennegen. The SED is typical for
  the sample, showing a large and broad IR-excess. A total reddening
  of E(B-V)=1.1 was deduced by minimizing the difference between the
  dereddened broadband fluxes (diamonds) and the appropriate model 
  atmosphere (full black line). The infrared excess was not considered
  in this minimalisation. The Spitzer infrared spectrum \citep{gielen08} is also shown.}
\end{figure}

A typical spectral energy distribution is shown in Fig.~\ref{fig:sed}.
The stellar flux reprocessed into the infrared is about 50 \% for this star.
The interested reader is referred to \citet{deruyter06} for the SEDs
of the complete sample.  In all objects of the sample, the dust excess
starts at, or very near to, dust sublimation temperature and this irrespective
of the spectral type of the central star. With a typical luminosity
of a post-AGB star, this sublimation temperature edge is at a distance smaller
than \mbox{10~AU} from the central source. We therefore inferred \citep[e.g.][]{vanwinckel03} that part 
of the dust must be gravitationally bound: any typical AGB outflow velocity 
would bring the dust to cooler regions within years. This is much shorter
than the likely evolutionary timescale of the central object.  We
argued that the same inner geometry, found in the resolved system
HD\,44179 \citep{menshchikov02,cohen04,bujarrabal05}, applies to the whole 
sample: the objects seem to be surrounded by Keplerian discs of dust.

Our recent interferometric studies confirm the very compact
nature of the circumstellar material \citep{deroo06,deroo07}. Our
Spitzer and ground-based N-band spectroscopic data show that dust grain 
processing is strong, with oxygen-rich dust dominated by, considerably larger, 
crystalline silicates \citep{gielen08}. Sub-mm bolometric data, available 
for a few sources only, show the presence of large grains in the circumstellar 
environment \citep{deruyter05}. These large grains have relatively small dust 
settling times, probably causing the disc to be inhomogeneous, consisting of 
small hot grains in the surface layer of the disc and a cooler midplane of 
mainly large grains \citep{gielen07}.

The actual structure of those discs, let alone their formation,
stability and evolution are not well understood.
We therefore started a very extensive multi-wavelength study of those systems.
In this contribution we report upon our detailed radial velocity
monitoring programme on the 1.2m Swiss Euler telescope. The
complementary photometric monitoring was performed at SAAO and on the
1.2m Flemish Mercator telescope. The main aim of this programme is to
investigate the link between binarity and the presence of
gravitationally bound material and to gain insight into the final
evolution of what appears to be a significant population of binary stars.

We organised the paper as follows:
In the next section we introduce the sample stars as well as the specific
criteria which were used to define the sub selection of objects discussed 
in this paper. In Section \ref{sect:data} we sketch the data gathering 
procedure as well as the reduction methods. In Section \ref{sect:radvel} 
we focus on the radial velocity determination and in Section \ref{sect:stars} 
we report on the analyses of the sample stars individually. 
In Section \ref{sect:discussion} we discuss the binary frequency of the sample
and the analyses of the orbital elements. We place our results in a broader 
context of stellar evolution.

\section{Sample}\label{sect:sample}

The total sample resulting from our specific selection criteria is 
described in detail in \citet{deruyter06}.

In this paper we restrict ourselves to the stars in our radial
velocity programme with spectral type F, five in total, which show only
small amplitude photometric variability with a peak-to-peak amplitude
of up to 0.25 magnitudes in the V-band and which are accessible from
the southern hemisphere. Additionally, we included HD~131356 (\entra), 
which was recognised as of RV Tauri type by \citet{pel76} and found to have 
a large excess at L by \citet{lloydevans85}, despite the higher peak-to-peak 
pulsational amplitude. The main reason for its inclusion is that, thanks to 
the longer time baseline of our velocity monitoring, we could confirm the 
binary nature and update the period estimate of \citet{vanwinckel99}. It 
shares the early F spectral type of the other stars discussed here, and in 
the 2001/2 season the V amplitude was only 0.3 magnitudes. 

The spectral types are from objective prism work \citep{houk76, macconnell76, 
kwok97} in the case of \hd and \entra. The four remaining stars, 
which were all initially selected for their large infrared excesses 
\citep{lloydevans99}, were classified by Lloyd Evans on blue 
light spectra of resolution 7$\,$\ang, using only the higher Balmer lines as 
being relatively unaffected by emission, since intrinsic metal deficiency and 
subsequent depletion weaken the metallic lines and vitiate the usual 
classification criteria. Luminosity class Ib was 
assumed as being broadly consistent with both the sharpness of the hydrogen 
lines and with the similarity of the other properties of these stars to 
Type II Cepheids. Published spectral types of B7III for IRAS 19125+0343 
\citep{kwok97} and B9Ib or B1III for IRAS 19157-0247 \citep{kwok97, 
parthasarathy00b, gauba03} may be discounted, as red spectra show strong 
lines of the CaII triplet, which have an excited lower state and so do not 
appear as interstellar absorption lines, whereas H and K lines of 
interstellar origin may be quite strong.
 
Cooler objects of the whole sample of \citet{deruyter06} lie in the
Population II Cepheid instability strip and have photospheric
pulsations of much larger amplitude. These stars have typical hydrogen 
line spectral types of G0 \citep{lloydevans99}. In many of those, the 
pulsations show substantial cycle-to-cycle variability.  These pulsations 
often show the defining alternating deep and shallow minima of RV\,Tauri 
stars. The detection of orbital motion in those objects is very difficult, 
because of confusion with the radial velocity variations induced by the 
photospheric pulsations \citep{maas02,vanwinckel98,vanwinckel99}.

In Table~\ref{tab:sample} the programme stars for this contribution are
listed with their spectral type, the equatorial coordinates, the
galactic coordinates and the visual mean magnitude. 

\begin{table*}
\caption{\label{tab:sample}General characteristics of the sample with the
  IRAS names, other identifications, the spectral type, the
equatorial coordinates, ($\alpha_{2000}$,  $\delta_{2000}$), the 
galactic coordinates, (l, b) and the mean visual magnitudes.}
\begin{center}
\begin{tabular}{lllrrrrrrrrr} \hline\hline\rule[0mm]{0mm}{3mm}
IRAS  & HD   & sp. type & \multicolumn{1}{c}{$\alpha_{2000}$} & 
\multicolumn{1}{c}{$\delta_{2000}$} & \multicolumn{1}{c}{l} & 
\multicolumn{1}{c}{b}   & \multicolumn{1}{c}{m(v)} & T$_{\rm eff}$(K)  & 
$\log$\,g & [Fe/H] & reference \\ 
\hline\hline\rule[0mm]{0mm}{3mm}
08544$-$4431   & V390 Vel   & F3   & 08 56 14.18  & $-$44 43 10.7  &
 265.51 & +0.39 &  9.13 & 7250 & 1.5 & $-$0.5 & \citet{maas05}  \\
12222$-$4652   & HD\,108015   & F3Ib & 12 24 53.50 & $-$47 09 07.5 & 
298.25 & +15.48&  7.95 & 7000 & 1.5 & $-$0.1 & \citet{vanwinckel97}\\
               & NSV 5601     &      &             &                &
       &       &       &      &     &        &     \\
14524$-$6838   & HD\,131356    & F2Ib & 14 57 00.69 & $-$68 50 22.9 &  
313.90& -8.68 &   8.77 & 6000 & 1.0 & $-$0.7 & \citet{vanwinckel97} \\
             & EN TrA        &      &             &                 &
       &       &        &     &    &         &    \\
15469$-$5311   &            & F3   & 15 50 43.80  & $-$53 20 43.3  & 
327.82 & +0.63 &   10.56 & 7500 & 1.5 & 0.0 & \citet{maas05} \\
19125$+$0343   &            & F2   & 19 15 01.17  & $+$03 48 42.7  &  
39.02 & -3.49 &   10.16 & 7750 & 1.0 & $-$0.5 & \citet{maas05}\\
19157$-$0247   &            & F3   & 19 18 22.71  & $-$02 42 10.8  &  
33.59 & -7.22 &   10.70 & 7750 & 1.0 & $-$0.4 & \citet{maas05} \\ \hline
\end{tabular}
\end{center}
\end{table*}

\section{Data and Data reduction}\label{sect:data}

In the framework of the cooperation between the Geneva Observatory and
the Institute of Astronomy in Leuven (Belgium), the twin 1.2m
telescopes Euler (La Silla) and Mercator (La Palma) were constructed.
Within this framework, the observers of Leuven have regular access to
the Swiss Euler telescope operated by the Geneva observatory and
the observers of the Geneva Observatory have regular access to the
Flemish Mercator telescope operated by the Institute in Leuven. During the
Leuven telescope time on Euler, typically 3 runs of 10-14 days every
semester, the programme stars were monitored in radial velocity with
the spectrograph CORALIE \citep{queloz99}. We started this campaign at
the end of November 1998.  
In Table~\ref{tab:dataradvel} we list the main characteristics of the 
observational data.

The CORALIE spectrograph is a fibre-fed spectrograph which samples the
visual spectrum in 68 spectral orders ranging from 388 nm to 682
nm. The resolving power ($\lambda/\Delta \lambda$) is 50000 at 500 nm,
with a 3 pixel sampling.  The fibre diameter is 2\,arcsec on the sky.  We
used the on-line data-reduction system \citep{queloz99} which includes
all standard steps in echelle reduction.

Older radial velocity data of \hd and \entra are also included. These
data were obtained with the CES spectrograph mounted on the CAT
telescope at La Silla (ESO) as well as with the CORAVEL instrument
mounted on the 1.5m Danish telescope also on La Silla.  CORAVEL is a
spectrophotometer by which the radial velocity was obtained by
cross-correlation of the stellar spectrum with a hardware mask built
from the spectrum of the K2~III star Arcturus \citep{baranne79}. \entra 
has stronger lines than the other stars observed, so that clear
correlation profiles are obtained from these data. The
zero-point of the system was regularly assessed during the night by
measuring CORAVEL radial velocity standards. These calibrations were
performed at intervals of about 1-2 hours. All the CORAVEL observations, 
including ours, are held in a database that is maintained by the owners 
of the CORAVELs in Geneva.

The stars accessible from La Palma were included in the long-term
photometric monitoring programme performed with the 1.2m Mercator
telescope using the refurbished Geneva 7-band dual-beam photometer
\citep{raskin04}. This resulted in 47 high quality absolute
photometric measurements for \eennegen but only 6 for \eennegenb. The
seven bands are measured quasi simultaneously using a filter wheel
that cycles through all seven filters four times per second
alternating as well between the sky and the object channel \citep{golay80}. 
The same instrument was previously attached to the now decommissioned 0.7m
Swiss telescope at La Silla. During that period, we obtained many
measurements for the Southern stars, \hd and \entra.

The photometric observations in the Cousins UBVRI system were obtained
from SAAO with the Modular Photometer on the 0.5m telescope and
subsequently with a very similar photometer on the 0.75m Automatic
Photoelectric Telescope \citep{martinez02}. These observations were
made with reference to the Cousins standards in the E-regions
\citep{menzies89}. The data are contained in Table S1, except for
\entra for which the observations were made by \citet{berdnikov95} and
\citet{caldwell01}. 

Since such monitoring programmes require the 
dedicated efforts of many people, the observers responsible for the
data acquisition for this project are included as co-authors in 
alphabetical order.

We complemented our own observations with the V-band data available
from the All Sky Automated Survey \citep{pojmanski02}. These data were
obtained with the ASAS-3 configuration using the wide field
(8.8$^{\circ}$x8.8$^{\circ}$) CCD camera with a sampling of
$\sim$15''/pix.  The critical PSF sampling is not reached and the
pipe-line reduction approximates aperture photometry with 5 circular
apertures from 2 to 6 pixels in diameter. The ASAS guidelines foresee
a preferred aperture for every magnitude bin and we limited our
analysis to measurements with quality flag A, according to the less 
rigorous classification applied to the re-reduced data on the ASAS website.
The recommended apertures were used, except that for \eennegen we used 
aperture 4, MAG-3, rather than aperture 1, MAG-2, to ensure full inclusion of the 
M star at an angular distance of 9.9\,arcsec. We applied corrections for the 
small offsets between the magnitudes of the same star observed with different 
cameras and for the offsets between the original and later reductions of 
early observations, many of which were not included in the re-reduced data 
on the ASAS website after recovery from the data storage crash. The 
photometric zero-point is tied to the Hipparcos catalogue, as in every 
frame a few hundred Hipparcos stars are located. These were used in the 
zero-point definition \citep{pojmanski02}. Near simultaneity of some of 
our observations with ASAS data enabled determination of zero point 
differences and error estimates. The offsets (ASAS-other), standard
deviation (s.d.) of the
single observations, the number of data points and the source of comparison 
data are: \nulacht, +0.043\,mag, 0.010\,mag, 14, 
SAAO; \hd, +0.009, 0.015, 44, SAAO; \eennegen, -0.057, 0.022, 18, P7; 
\eennegenb, +0.043, 0.023, 3, P7. These differences have been allowed for 
in combining the relevant datasets. Additionally, from non-overlapping 
datasets, we have offsets for \eenvijf, -0.030, SAAO; \eennegen, +0.045, 
SAAO; \eennegenb, +0.070, SAAO. The SAAO magnitudes of \eennegen are 
0.09\,mag brighter than the P7 observations, because of the inclusion of 
the M star in the aperture used at SAAO. These last values will include any 
actual changes in brightness between the two epochs. 

The shift between SAAO and P7 for \eennegen is attributable to the inclusion 
or exclusion of the nearby M type star. The reason for offsets for other stars 
is not known. The s.d. obtained by direct comparison of photometries 
are smaller than the errors ascribed to individual observations in ASAS, which 
must be more precise than was previously recognised. Intercomparison of three 
independent subsets within the ASAS data for \nulacht gives values in the 
range 0.011 to 0.024\,mag, confirming the typical values above.    

The ASAS data proved very valuable to compensate for the poorer sampling of 
our own multi-colour measurements in the search for the long periods expected. 
ASAS data were used by \citet{kiss07} in their study of the pulsations of 
these and similar stars, but the continuation of the ASAS project has doubled 
the timespan of the observations, which now extend from 2001 or 2002 to the 
end of the 2008 season. Moreover, the re-reduction of much lost ASAS data has 
filled the gap around JD\,245\,3500 seen in the plots by \citet{kiss07} for 
these stars. The more extensive data facilitate the characterisation of the 
noisy small amplitude oscillations of these stars.

In  Table~\ref{tab:dataphot} 
we list the main characteristics of the raw photometric data sets for all objects.

\begin{table}
\caption{\label{tab:dataradvel}General characteristics of the
spectral dataset obtained for every object. For \hd and \entra
the CORALIE radial velocity data were complemented with velocities 
obtained by earlier high-resolution spectroscopic measurements. N$_{tot}$ 
is the number of observations, $N_{used}$ the number used after our quality 
test (see text), and $\Delta \rm HJD$ is the time range of the observations. 
The total peak-to-peak radial velocity spread, $\Delta$V$_{\rm rad}$ (\kms) 
and $\Delta V$, the visual photometric peak-to-peak variation, are also shown.}
\vspace{1.ex}
\begin{center}
\begin{tabular}{lrrcrr} \hline\hline \rule[0mm]{0mm}{3mm}
Object   & N$_{tot}$ & N$_{used}$ & $\Delta \rm HJD (24+)$ & $\Delta$V$_{\rm rad}$ &
$\Delta V$ \\
IRAS    &    &    &                        &      \kms                 & \\
\hline
08544$-$4431   & 181 & 161 & 51147-54487 & 22.1 & 0.25 \\
12222$-$4652   & 83  &  82 & 45397-53833 & 13.5 & 0.23 \\
14524$-$6838   & 75  &  63 & 48313-53904 & 45.7 & 0.81 \\
15469$-$5311   & 175 & 161 & 51278-54680 & 28.2 & 0.25 \\
19125$+$0343   & 103 &  90 & 51278-54331 & 29.6 & 0.19 \\ 
19157$-$0247   & 137 & 111 & 51279-54680 & 24.2 & 0.23 \\ \hline
\end{tabular}
\end{center}
\end{table}

\begin{table*}
\caption{\label{tab:dataphot}General characteristics of the
photometric datasets obtained for every object. The number of
measurements in the different photometric systems is given. P7 is the 7-band
Geneva photometric system, SAAO is the Cousins UBVRI-system, and the
ASAS monitoring is obtained in the V-band only.  In brackets the
total time coverage in HJD (24+).} 
\begin{center}
\begin{tabular}{lccc} \hline\hline\rule[0mm]{0mm}{3mm}
Object       & P7  & SAAO  & ASAS \\
\hline
\nulacht   &                   &   91 (50858-52060) & 830 (51868-54843) \\
\hd        & 120 (47258-47824) &  166 (51625-53195) & 576 (52441-54711) \\
\entra     &  61 (47608-49853) &   58 (43979-49564) & 633 (51905-54743) \\
\eenvijf   &                   &   49 (48803-51712) & 550 (51925-54743) \\
\eennegen  &  47 (52415-53606) &   30 (49485-51819) & 362 (52442-54772) \\
\eennegenb &   6 (52446-52561) &   50 (48404-51819) & 386 (51979-54777) \\ \hline
\end{tabular}
\end{center}
\end{table*}

\section{Radial velocity determination}\label{sect:radvel}

To obtain accurate radial velocity information based on low S/N
spectra, we determined cross-correlation profiles for all individual 
measurements. The standard spectral mask includes a large set of well
defined spectral lines for the given spectral type.
Our chemical analysis, based on high signal-to-noise optical spectra, 
showed that for most stars several elements, including iron, 
are under-abundant \citep{maas05}. By eliminating, from the standard 
spectral mask, those lines not detected in our high-quality spectrum, 
we determined an appropriate spectral mask for every individual star.
Moreover, we selected only single lines with clear symmetric profiles.

\begin{figure}
\resizebox{\hsize}{!}{\includegraphics{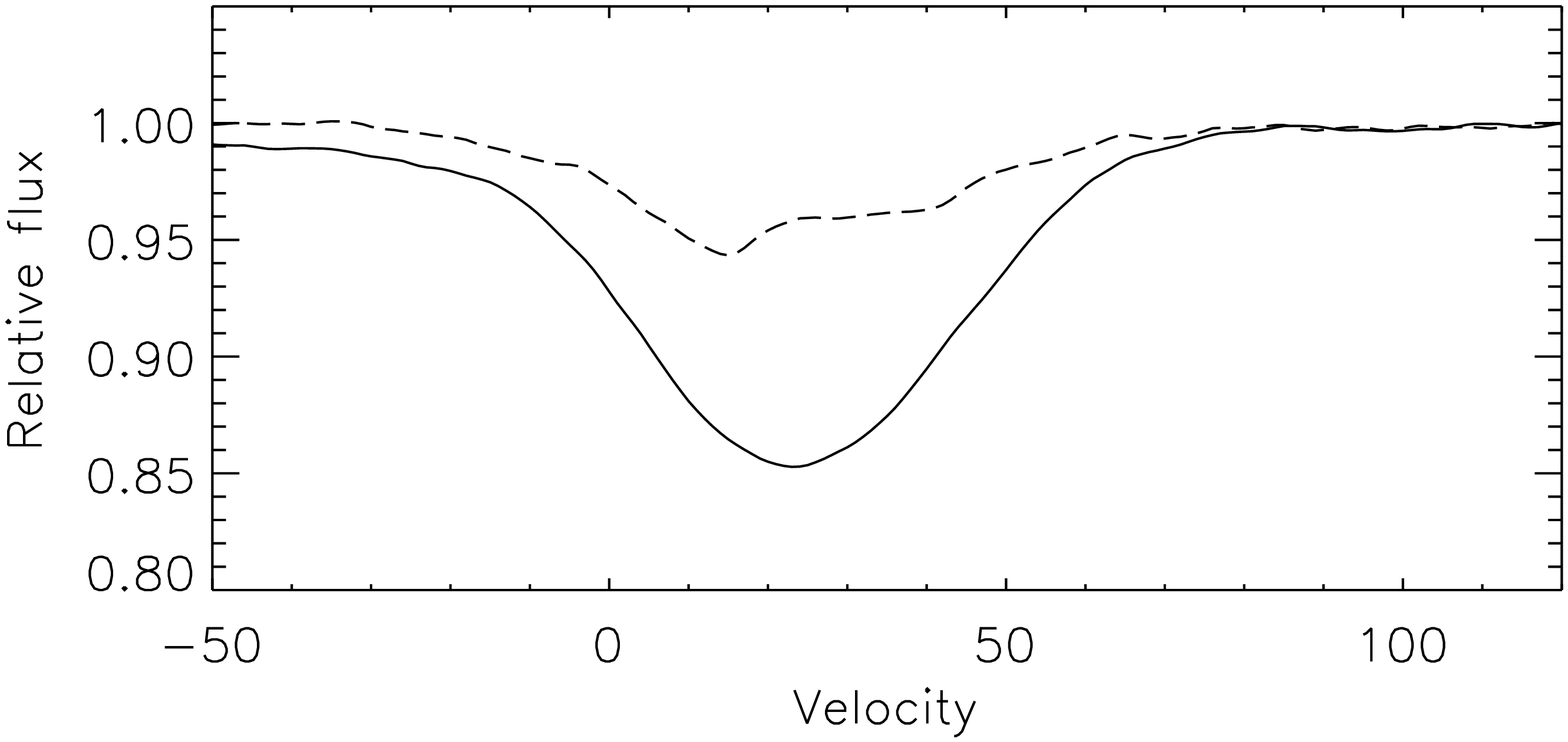}}
\caption{\label{fig:maskerverg} Comparison between cross-correlation
  profiles obtained with a standard spectral mask (dashed line) and
  with a software mask adapted to the individual spectral
  characteristics of \eennegenb (full line).}
\end{figure}

In Fig.~\ref{fig:maskerverg} we compare the cross-correlation profile
obtained by the standard mask with the one obtained by the specific
mask, constructed on the basis of the high signal-to-noise spectrum
for \eennegenb.  No accurate velocity can be derived on the basis of
the standard mask generated profile, while a Gaussian fit of the
profile, obtained by the specified mask for IRAS\,19157-0247, gives a
good estimate of the radial velocity.

The radial velocity is computed using the mean of a 50-point bisector
of the cross-correlation function. An initial Gaussian fit was used to
determine the bisector points which are defined on an equidistant
velocity grid centred on the velocity minimum of the Gaussian fit.  The
grid spans twice the Gaussian width ($\sigma$).  The internal
uncertainty of a radial velocity measurement is estimated by the rms
of these 50 bisector values. We eliminated those cross-correlation
profiles for which the rms is larger than 1 \kms. Several spectra were
obtained under very poor weather conditions and/or failed due to
guiding failure. These measurements were eliminated in this step of the
analysis. Despite the fact that the objects do not have a large
pulsational amplitude, some objects do show asymmetries in the 
cross-correlation profiles at particular phases in the pulsation cycle.
These profiles are also eliminated at this point since we focus on the
detection of orbital motion. The number of radial velocity
measurements which passed our critical quality assessment test is give
in Table~\ref{tab:dataradvel}. 
The individual datapoints of all objects 
are given in Table 4 which is available at CDS. \addtocounter{table}{1}

\section{Variability study of individual stars}\label{sect:stars}

The main goal of our programme was to look for evidence of binary
motion in the radial velocity data and, if found, determine the
orbital elements. Although for this paper we selected objects
with a small pulsational amplitude in V, they all show some
photometric variability which could be present in the radial velocity
data as well. 

Luminous stars on the blue side of the population II Cepheid instability strip are,
however, far from regular mono-periodic pulsators and often show a very
complex pulsational behaviour. One example is HD\,56126
\citep{barthes00, fokin01} where the main pulsation mode is shown to
generate shocks provoking a complex asynchronous motion in the outer
layers. Another example is HR\,7671 (HD\,190390) where the main period
of 28.6 days is accompanied by a beating behaviour either due to a
stable pulsation triplet or due to an unstable main period. Consequently
the description of the pulsation in radial velocity, modelled as a simple
harmonic, fails to reproduce the complete dataset which shows a 
peak-to-peak difference of 8.2\,\kms \citep{reyniers05}, as well as a 
corresponding light amplitude of 0.36 mag. In neither object is there 
evidence for binary motion.  Since the programme stars of this project are 
also on the hot side of the instability strip, we
investigated the photometric dataset in parallel with the radial
velocity data as to confront radial velocity variability with
photometric behaviour.

For the variability study we used phase dispersion minimization 
{\rm PDM} \citep{stellingwerf78} and, for the photometric data, the 
Lomb-Scargle method to search for periodicities. The orbital solutions were 
then computed using an altered version of the Fortran code {\rm VCURVE}
\citep{bertiau69} as well as the minimisation code {\rm FOTEL} 
\citep{hadrava04b}. The weights of the individual velocity points were 
based on the rms values of the bisector analyses. All errors quoted in 
the individual orbital elements are rms errors of these elements calculated 
from the covariance matrix. Plots of the photometry of single seasons were used to 
look at the individual cycles in some cases.  The individual
photometric datapoints of all stars
are given in Table 5  and Table 6 available at CDS. 
\addtocounter{table}{2}

\subsection{\nulacht}

\begin{figure}
\resizebox{\hsize}{!}{\includegraphics{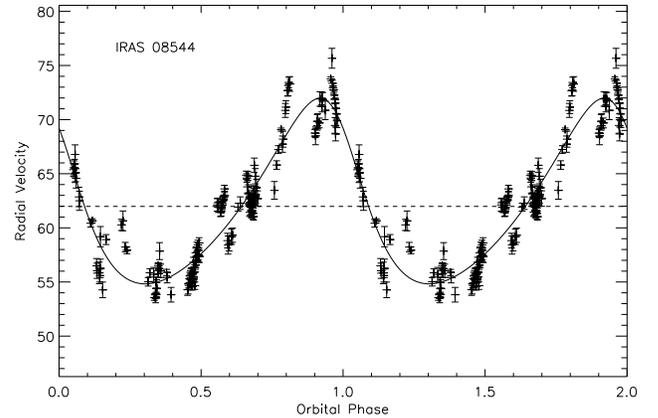}}
\caption{\label{fig:08544orbit} The radial velocities folded on
the orbital period of 508 days for IRAS\,08544-4431. The errorbars are
defined as the standard deviation on the 50-point bisector of the
cross-correlation profile. The system velocity is given by a dashed line.}
\end{figure}

We have already shown that this object is a spectroscopic binary in
\citet{maas03}. We continued our monitoring of this object to refine 
the orbital elements. We have gathered a total of 161 good
radial velocity measurements sampling more than 6 orbital
periods. The resulting radial velocity variations are shown in 
Fig.~\ref{fig:08544orbit}. An orbital period of 508 days and an 
eccentricity of 0.25 were found. All orbital elements
are listed in Table~\ref{tab:orboverview}.

Our previous pulsational analysis of photometry from SAAO \citep{maas03} 
has shown that low-amplitude pulsations are present, but that the cycle-to-cycle
variability is large. Periods of 72.3 and 89.9\,d were reported. \citet{kiss07} 
confirmed the shorter period and obtained two new periods, 68.9\,d and 133\,d, 
from ASAS data which extended to JD\,245\,3900, with a break 245\,3200 - 
245\,3750.  

\begin{figure}
\resizebox{\hsize}{!}{\includegraphics{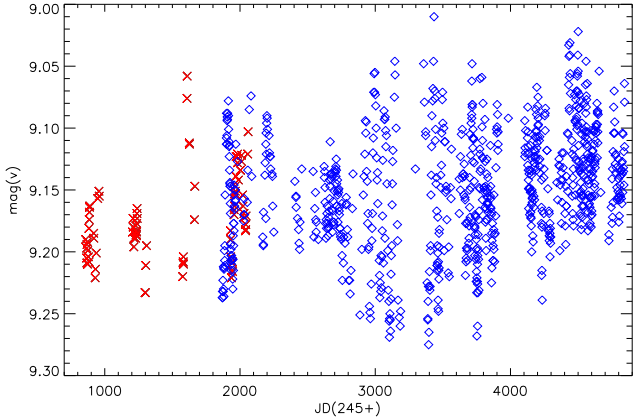}}
\caption{\label{fig:08544phot} The combined datasets of the V-band
  photometric measurements of \nulacht.  The x-symbols are the 
SAAO data and the diamonds are the ASAS measurements.}
\end{figure}

We combined the SAAO and the best current ASAS dataset (see Table~\ref{tab:dataphot}) 
and analysed the total V-band time-series. In Fig.~\ref{fig:08544phot} the total
dataset is shown and a beating pattern with a clear amplitude modulation is 
present, as well as a slow brightening. The dominant period of 
71.7 $\pm$ 0.4\,d gives in a harmonic fit an amplitude of 0.03\,mag and a 
variance reduction of 21 \%.

\begin{figure}
\resizebox{\hsize}{!}{\rotatebox{90}{\includegraphics{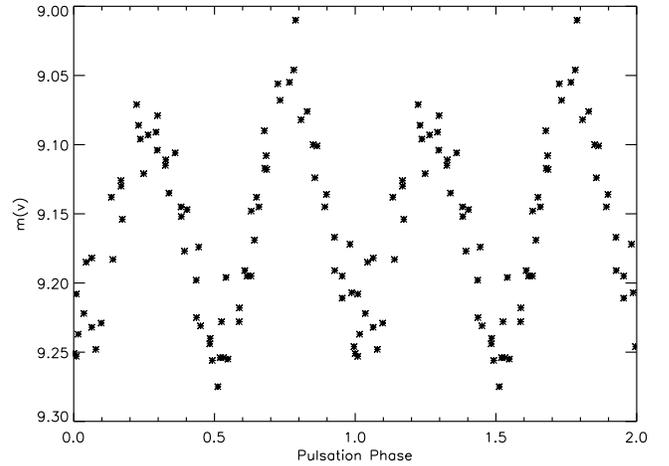}}}
\caption{\label{fig:08544puls}
The V magnitudes for \nulacht plotted against the double
period, 143.6 days.  Data of period JD
245\,2887-245\,3397 when \nulacht showed a large amplitude. }
\end{figure}

The amplitude was greatest, at 0.20\,mag, in the interval JD\,245\,2850 to 
245\,3500.  Fig.~\ref{fig:08544puls} shows these data, phased on the double period 143.6\,d.
The alternate minima are of different depth and the alternate maxima 
are of unequal height, though the differences are quite small. The
behaviour was maintained for almost
2 years and is reminiscent of the typical RV\,Tauri lightcurves, albeit
with a much smaller amplitude.
Much of the small scatter may be attributed to the observational error of about 
0.02\,mag. Corresponding plots 
for the two intervals of small amplitude, represented by SAAO data from 
JD\,245\,0850 to 245\,1500 and ASAS data from JD\,245\,2400 to 245\,2750 show 
amplitudes of 0.08\,mag and 0.06\,mag respectively, while the curves are less 
well defined. The separation of these intervals, at least 1400\,d, is less than 
that to be expected from a beat between the two principal periods, 2800\,d, 
as suggested by \citet{kiss07}, but the next interval of small amplitude had not 
occurred by early 2009, JD\,245\,4840, and may yet fit the prediction.

There is no evidence for the orbital period in the ASAS photometry 
\citep{kiss07} nor in our extended multicolour photometry. Given the 
complex atmospheric pulsations, we did not attempt to clean the low 
amplitude irregular pulsational variability from the radial velocity 
dataset.

\subsection{\hd}

\begin{figure}
\resizebox{\hsize}{!}{\includegraphics{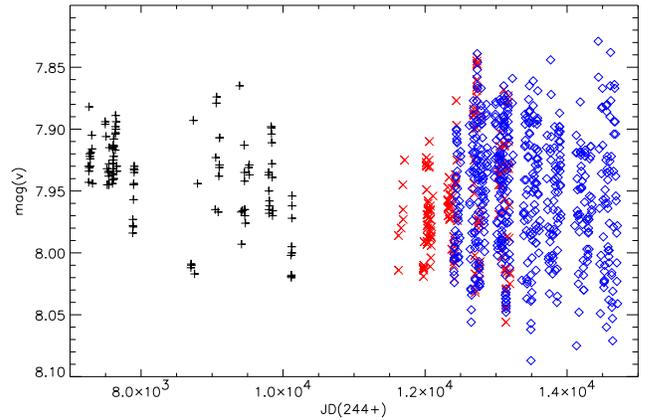}}
\caption{\label{fig:108015phot} The combined datasets of the V-band 
photometric measurements of \hd.  The + are the Geneva 
measurements, the x-symbols are the SAAO data and the diamonds are the 
ASAS measurements.}
\end{figure}

\hd also shows a complex pulsational pattern with a significant
amplitude modulation and with a smaller amplitude at earlier times. 
(see Fig.~\ref{fig:108015phot}). ASAS data for the period before 
JD\,245\,2300 were highly erratic, in contrast to contemporaneous SAAO 
data, and have been rejected. Our Geneva photometry is poorly sampled 
over many years, while the SAAO and ASAS photometry combine well to give 
a larger dataset. We find a dominant period of 60.5 $\pm$ 0.11\,d and a second 
period of 55.3 $\pm$ 0.2\,d, in agreement with \citep{kiss07}. A combined 
harmonic fit gives a fractional variance reduction of 52 \%. Plots of the 
V magnitude against time show continuous variability in the amplitudes and 
lengths of successive cycles, from 0.11 to 0.20\,mag and 48 to 76 days. 
The longer cycles tend to have amplitudes which are below average. The PDM 
analysis also yields an additional period of 120.6 $\pm$ 1.0\,d, or twice the 
principal period.  \hd is clearly another star with a complex dynamical 
photosphere which complicates the interpretation of radial velocity variations.

\begin{figure}
\resizebox{\hsize}{!}{\includegraphics{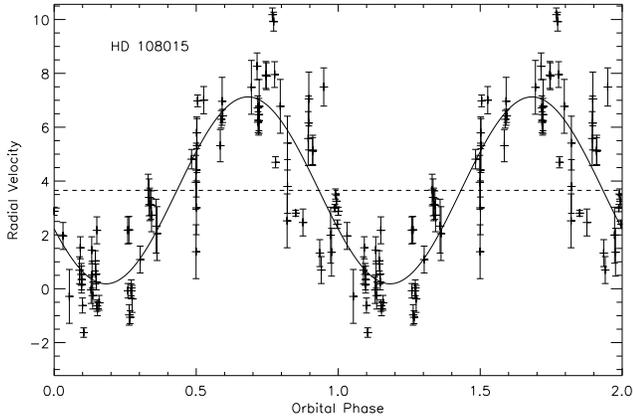}}
\caption{\label{fig:108015orbit} The radial velocity variations of \hd,
after correction of the variations induced by the pulsations. The data
are folded on the orbital period of 914 days. The full line is a
circular model for the orbit.}
\end{figure}

The total radial velocity dataset of \hd includes older data, measured on 
high-quality spectra obtained to study the chemical composition
\citep{vanwinckel97}. Our total timespan is therefore more 
than 23 years. The dataset gives a significant initial period of 
917 $\pm$ 4 days, which we interpret as due to orbital motion.
A circular orbit gives a fractional variance reduction of 65 \%. After
cleaning of this first circular model, we retrieve the dominant photometric
period of 60.5\,d in our $\theta$-statistic with an amplitude of 1.4\,\kms.
Our final orbital solution (see Fig.~\ref{fig:108015orbit}) is then obtained 
after cleaning the original dataset with the harmonic fit of the dominant 
photospheric pulsation mode. The small eccentricity is not significantly
different from zero, so our final solution gives a circular orbit
of 914 $\pm$ 4 days. The  fractional variance reduction is 75\%.
The orbital period is not present, either in the magnitudes or in the colours.

\subsection{\entra}

HD\,131356 (EN\,TrA) was discovered to be variable by Miss Leavitt 
\citep{pickering07}. The binary nature and preliminary orbital elements were 
given in \citet{vanwinckel99}. We have since accumulated more data using
the CORALIE spectrograph in order to improve the orbital solution.

The object is catalogued in the GCVS as a small amplitude Cepheid with
a published pulsation period of 36.54\,d \citep{grayzeck78}, based
on a limited dataset covering part of two pulsational cycles, with
poor coverage in phase. \citet{pel76} obtained a continuous
time-series of 44 measurements on the Walraven photometric system. The
object showed at that time (JD\,2440990-2441171) a substantial V-band
amplitude, with a maximum of V=8.55, a deep minimum of 9.2 and a
shallow minimum of 8.8. The Hipparcos catalogue (HIP 73152) list a
period of 37.10 days \citep{perryman97}.

Our Geneva photometry is poorly sampled and no
periodicity could be determined. Two sets of observations obtained at
SAAO on the UBVRI system are available, comprising 42 observations
(JD\,2443979-2445155) by \citet{caldwell01} and 16 observations
(JD\,2449522-2449564) by \citet {berdnikov95}. Both indicate a range
of V=8.5-9.05, with little difference between deep and shallow
minima. A short dataset by \citet{eggen86} gives V=8.55-8.96, without
full coverage of the extrema. The first season of ASAS data, 2001-2002, 
shows a variable amplitude of up to 0.3\,mag, similar to the other stars 
(see Fig.~\ref{fig:entralightcurve}).   

\begin{figure}
\resizebox{\hsize}{!}{\rotatebox{90}{\includegraphics{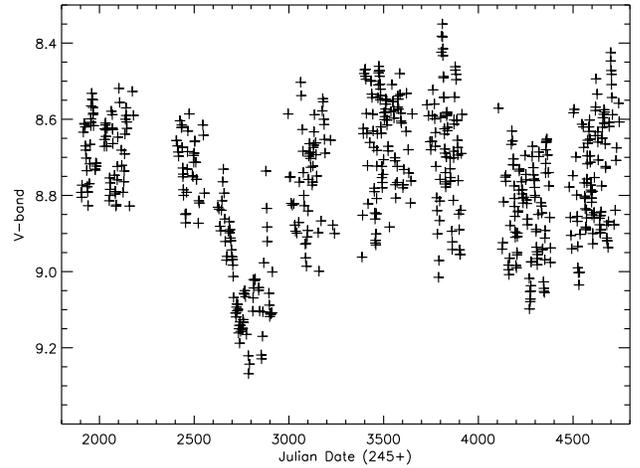}}}
\caption{\label{fig:entraasas} The V-band photometric dataset of \entra obtained
by the ASAS project \citep{pojmanski02}.}
\end{figure}

\entra fades by some 0.5 mag near JD\,245\,2800 and again, by about
0.1\,mag, near JD\,245\,4300 (see Fig.~\ref{fig:entraasas}). Outside
of these intervals, the mean magnitude is fairly constant but the
amplitude is variable. Period analysis gives a single period P = 37.04
$\pm$ 0.03\,d and a double period 74.1 $\pm$ 0.1\,d. Aliases
correspond to the orbital period and the calendar year. The magnitudes
at maximum and minimum light are both very variable from cycle to
cycle and from year to year. The deep and shallow minima, and bright
and faint maxima, generally retain the same relative phasing within a
given season, but they sometimes interchange in successive years.
These phenomena are well known in RV Tauri stars.

\begin{figure}
\resizebox{\hsize}{!}{\rotatebox{90}{\includegraphics{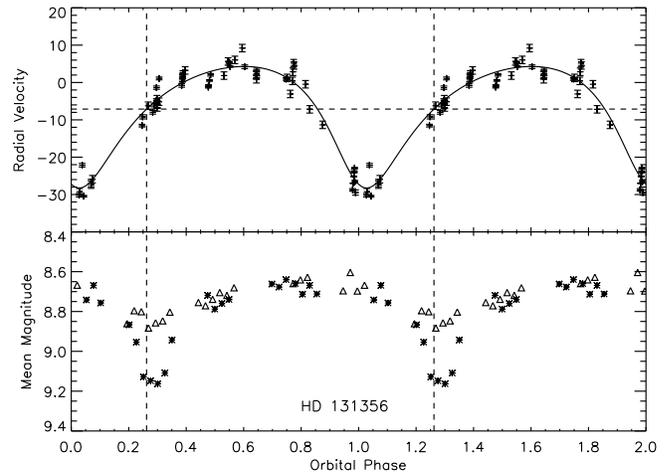}}}
\caption{\label{fig:entraorbit} The top panel shows the radial
  velocity variations of \entra folded onto the orbital period of
  1493. days. The full line represents the orbital model. The lower
  panel gives the mean magnitude determined on those pulsation cycles with more
  than six data points. The different symbols represent two different
  orbital cycles. Mean minimal light is measured when the object
  is at inferior conjunction but in the two different orbital cycles,
  the effect is not equally prominent.}
\end{figure}

\begin{figure}
\resizebox{\hsize}{!}{\rotatebox{90}{\includegraphics{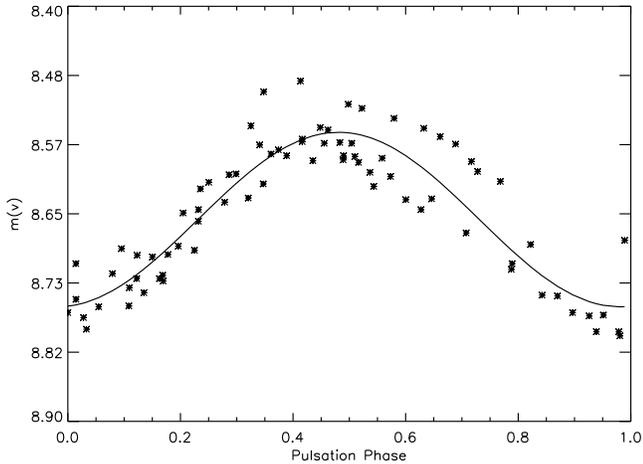}}}
\caption{\label{fig:entralightcurve} The V-band magnitudes of \entra, from the 
ASAS photometric monitoring project of the first season (JD\,2451903-2452173), 
folded on the pulsation period of 36.70 days found in the radial velocity data.}
\end{figure}

Thanks to the very long baseline of our radial velocity data, we could
sample about 3.7 orbital cycles. In a first iteration, 
after the removal of the orbital solution, the residuals showed a very
clear periodicity with a period of 36.70\,d, which we link to 
the pulsation timescale found in the photometry. A harmonic fit gave a
radial velocity amplitude of 4\,\kms and we cleaned the original data
with this mean pulsation model. The final solution gives an orbit with
a period of 1493 $\pm$ 10 days and a large
eccentricity of 0.31 $\pm$ 0.05. (see Fig.~\ref{fig:entraorbit}, 
Table~\ref{tab:orboverview}).

To investigate the secular variations in the mean magnitude we
selected only those pulsation cycles for which 6 or more data points
are available. We then cleaned for the pulsation amplitude and reduced
the pulsation cycle to the mean magnitude alone. This mean magnitude
is around 8.6 but there is a significant drop when the object 
enters the conjunction phase (see Fig.~\ref{fig:entraorbit}). 
The minimum (mean magnitude of 9.1) is
obtained exactly at inferior conjunction and \entra then slowly
recovers to its nominal mean values about 0.3 in orbital phase later. 
There are shallower fadings at the same orbital phase: three orbits 
earlier, near JD\,244\,8350 (P7 data) and JD\,245\,4300 (ASAS data). 
The corresponding orbital phase is not covered by any of the other datasets. 
The most natural explanation for the secular variability is 
that the star suffers from variable circumstellar reddening during orbital 
motion. The minimum is reached when the object suffers from a maximal 
reddening of circumstellar material in the line-of-sight, which would occur 
when the star is at inferior conjunction if there is a circumbinary disk.

\subsection{\eenvijf}

\eenvijf is another low-amplitude pulsator which is little studied in
the literature. The mean magnitude varies, with a fairly abrupt brightening 
of 0.045\,mag after JD\,245\,4000, as well as smaller variations on a time 
scale greater than 1000 days. We combined the ASAS and SAAO photometric V-band 
measurements for the pulsation study. The main period is 54.5 $\pm$ 0.3\,d, 
consistent with 54.4 $\pm$ 1\,d found by \citet{kiss07}, and a harmonic fit 
gives a fractional variance reduction of 21 \% with an amplitude of only 
0.028\,mag. The extended dataset did not confirm the second pulsation period 
of 49.1 days given by \citet{kiss07}; this might be a one year alias of the 
primary period. The cleaned data yield the next possible period of
56.6 $\pm$ 0.3 d. but the variance reduction is limited. Possible
other periods of 164 and 77 days may be
spurious as well, as they 
also appear in the windowed data.  Plots of magnitude against 
JD for 20 well observed cycles in four seasons give lengths of 32 to 77 days, 
with a mean of 54.0 days. The amplitude varies from 0.02\,mag to 0.15\,mag. 
The mean amplitude quoted above is an underestimate of the true variability, 
as the mean light curve is blurred by variability in both amplitude and 
length of period. The double period also emerges from the PDM analysis and a 
plot of V against 109.2 days shows a weak tendency towards alternate deep and 
shallow minima, as well as bright and faint maxima.

\begin{figure}
\resizebox{\hsize}{!}{\includegraphics{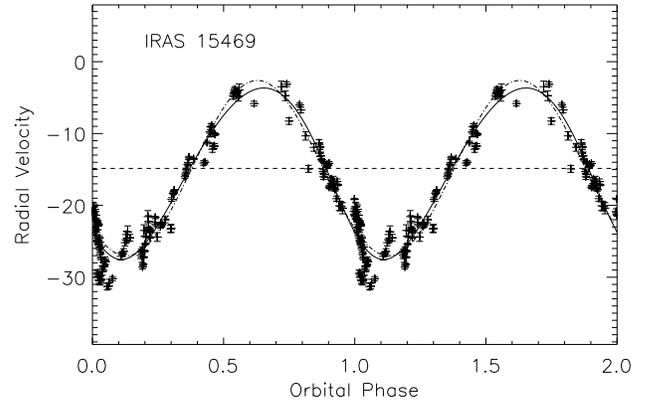}}
\caption{\label{fig:irasvijftien} The radial velocities of \eenvijf folded 
on the orbital period of 389.9 days. The full line is the orbital solution,
and the dash-dotted line is the best circular orbital curve.}
\end{figure}

We have 160 good radial velocity measurements for \eenvijf over a
time range of 3402 days. The orbit of 390 days is very well defined 
(see Fig.~\ref{fig:irasvijftien}). The classical
Lucy and Sweeney test \citep{lucy71} shows that the orbit has a small
but significant eccentricity of 0.09.  In Table~\ref{tab:orboverview} 
all other orbital parameters are given. There is an indication of the orbital 
period in the photometric data, for which \citet{kiss07} give a period of 
384 $\pm$ 50 days, but the phase diagram is not convincing. The residual 
velocity variations have a peak-to-peak amplitude of 10\,\kms (standard 
deviation of 2.1\,\kms) but there is no clear indication of the main 
pulsation period nor could we discover a clear periodicity in the residuals.

\subsection{\eennegen}

\eennegen is also a poorly studied object, despite the fact that it
combines rather bright IRAS fluxes with a significant visual flux of
m(v)=10.16. The LRS-IRAS spectrum is featureless \citep{kwok97} and
the source was not detected in the OH survey of \citet{lesqueren92}. 
The object was rediscovered by \citet{lloydevans99} in the search for 
new RV\,Tauri stars.

The photometric variability is small: the SAAO and P7 datasets give 
ranges of 0.11 and 0.07\,mag, respectively, in the V-band.  
\citet{kiss07} found a possible long period of 2300\,d; the more 
extensive ASAS data available now are consistent with this, with an 
amplitude of about 0.03\,mag, but instead a step change of 0.03\,mag 
fainter after JD 245\,3400 is a possibility. 

\begin{figure}
\begin{center}
\resizebox{\hsize}{!}{\includegraphics{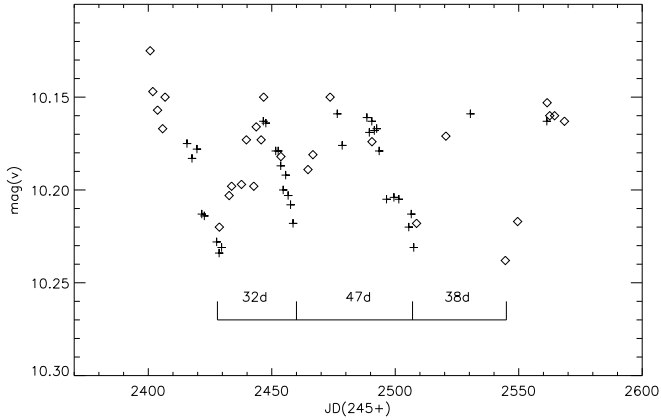}}
\caption{\label{fig:19125phot}
Detail of the lightcurve of \eennegen. Period changes occur on irregular
time intervals illustrating the strong cycle-to-cycle differences
observed in this object. P7 data are denoted by plus signs and ASAS data by diamonds.}
\end{center}
\end{figure}

Application of the period search algorithms to the complete dataset yielded 
several periods between 34 and 43 days: 42.3 and 38.9 days were the strongest. 
Phase plots against any of these periods showed much greater scatter
than can be attributed to observational errors. 
The light curves (Fig.~\ref{fig:19125phot}) for individual seasons were inspected, and the times 
between identifiable features, generally near the mean magnitude on the 
ascending or declining branches of the light curve, were estimated for 
successive cycles. There were insufficient data in 2006, and variations 
were small and/or erratic in 2004, 2005 and 2007. Estimates of the 
period could be made for 1994, from SAAO data, for 2002, from P7 and ASAS data, 
and for 2003 and 2008 from ASAS data. The values for 15 cycles ranged from 29 
to 51 days, with a mean of 38.2 days. On two occasions an incipient decline 
after maximum was reversed to give a second maximum and 
an unusually long interval between mid-decline points. This occurred in 2002, 
turning a tentative 32 day interval into 47 days, and in 2003, turning a 39 day 
interval into 51 days. The 519\,d orbital period is not present in 
the photometry.

\begin{figure}
\begin{center}
\resizebox{\hsize}{!}{\includegraphics{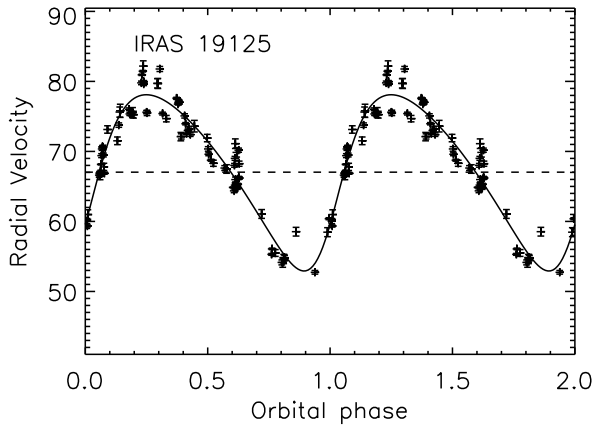}}
\caption{\label{fig:19125orbit}
The radial velocity data are folded on the orbital period of \eennegen, 
519.6 days. The full line is the orbital model.}
\end{center}
\end{figure}

We have 91 good radial velocity measurements over a time range of 3389
days.  We found a clear periodicity of 519 days and the peak-to-peak velocity
amplitude is no less than 29.5\,\kms. Given the period and amplitude,
we interpret this variability as due to orbital motion. We obtained orbital 
elements with a significant eccentricity of 0.26 $\pm$ 0.03.  We sampled in
total 6.5 orbital cycles. Fig.~\ref{fig:19125orbit} shows the radial
velocities folded on this period. The full line represents the best
Keplerian orbital solution (see Table~\ref{tab:orboverview}).

The strongest peak in both the PDM and Scargle periodograms for the residual 
velocity variations corresponds to 43 days, which leads to a variance reduction 
of 39 $\%$, and is fairly close to the mean of the cycle lengths found in the 
photometry. The orbital period is much shorter than the possible long 
photometric period \citep{kiss07} in the ASAS data.

\subsection{\eennegenb}

The SAAO and ASAS photometry show that \eennegenb varies with an amplitude 
of up to 0.15\,mag but there is no long term trend, although there is possible 
modulation in the amplitude. The ASAS photometry, whether taken as 
a whole or subdivided into shorter intervals, or supplemented by the 
SAAO and P7 data, consistently indicates periods of 22.5 to 22.8 days, 
but with low significance. This period does not give a satisfactory light 
curve for any of these datasets, even for the data obtained in a single 
season: there is only a weak tendency to a variation with phase in the 
derived period. Detailed examination of the time series for individual 
seasons shows great variability in both amplitude and length of successive 
cycles. The data are undersampled for this star, making delineation of the 
variability hard. \citet{kiss07} found an irregular variation, with a possible 
periodicity of 25 days.

We have 111 radial velocity measurements over a time range of 3397
days.  Both PDM and a Fourier analysis indicate that the 120 day
period is significant. This period in radial velocity is not present
in the photometry and we interpret this as the orbital period. The
non-pulsational origin of the 120 day period is corroborated by two
arguments. First, 120 days is a very long pulsational period for an
F-type post-AGB star, and in any case it is not found in the
photometric analysis.  Secondly, the peak-to-peak variations in V for
IRAS\,19157-0247 average 0.13\,mag at most, so it is very doubtful
that the radial velocity variations with a peak-to-peak amplitude of
24\,\kms are only due to pulsations. Interpreting the velocity curve
as due to pulsations of the photosphere, the integrated velocity would
yield a radius increase of about half the initial radius. Such a
pulsation would clearly not go unnoticed in the photometry. We
therefore interpret the velocity variations as due to orbital motion.
This is a very short orbital period with a high eccentricity which
puts strong constraints on the evolutionary history of the star !

\begin{figure}
\begin{center}
\resizebox{\hsize}{!}{\includegraphics{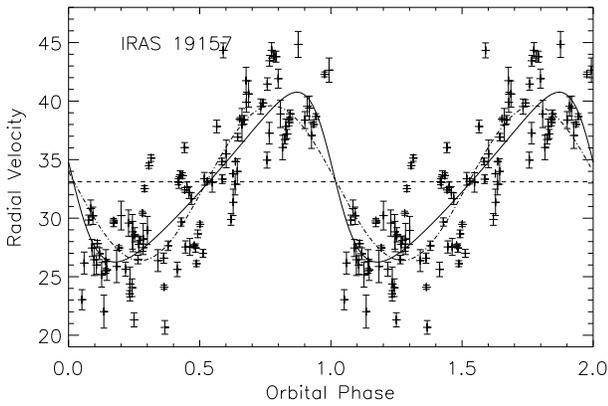}}
\caption{\label{fig:19157orbit} The radial velocities  of \eennegenb
  folded on the orbital period of 119.5 days. The full line is the
  best orbital model, the dashed-dotted line is the circular
  model. The horizontal line indicates the system velocity.} 
\end{center}
\end{figure}

Fig.~\ref{fig:19157orbit} shows the radial velocities and the
residuals folded on this period. We found an eccentricity of 0.31. The
other orbital elements are shown in Table~\ref{tab:orboverview}. The
non-zero eccentricity was tested with the classical Lucy and Sweeney
test \citep{lucy71}. The $\sum({\rm O} - {\rm C})^2$ is 14\% larger
for a circular fit. This, together with the many cycles sampled, shows 
that the eccentricity is significant. The periodograms for the residual 
velocities show several peaks between 9 and 20 days, probably related to 
the timescale of the pulsations. None of them lead to a significant 
variance reduction, however.

\begin{table*}
\caption{The orbital elements of IRAS\,08544-4431, IRAS\,12222-4652 (HD\,108015), 
EN TrA, IRAS\,15469-5311, IRAS\,19125+0343 and IRAS\,19157-0247 :
P, the orbital period in days, K, the velocity amplitude, e, the eccentricity, 
$\omega$, the angle in the orbital plane between the direction to the ascending 
node and the direction to the periastron, T$_{0}$, the time of periastron passage, 
$\gamma$, the system velocity, r.m.s. (O-C), the root mean square of the errors, 
$a_1 \sin i$ with a the semi major axis, F(M), the mass function in solar masses, 
N, the number of observations, $\Delta t$ (cycles), the number of cycles used 
and $\Delta V$, the visual photometric peak-to-peak variation,
are shown.}
\vspace{0.5ex}
\label{tab:orboverview}
\begin{center}
\begin{tabular}{lllllllllllll} \hline\hline\rule[0mm]{0mm}{3mm}
 & \multicolumn{2}{c}{IRAS\,08544} &  \multicolumn{2}{c}{IRAS\,12222} &  
\multicolumn{2}{c}{EN\,Tra} & \multicolumn{2}{c}{IRAS\,15469} &  
\multicolumn{2}{c}{IRAS\,19125} &  \multicolumn{2}{c}{IRAS\,19157} \\
 & & $\sigma$ & & $\sigma$ & & $\sigma$ & & $\sigma$ & & $\sigma$ & &
 $\sigma$ \\ \hline\rule[0mm]{0mm}{3mm}
Period (days) & 507.8  & 1.5 & 913.8  & 4.3 &
1493.  & 7. & 389.9     & 0.5  & 519.6  & 2   & 119.5   & 0.2\\
K (\kms)                           & 8.6    & 0.3 & 3.5    & 0.5 &
16.3  & 0.6 & 11.9      & 0.3  & 12.6    & 0.5 & 7.3     & 0.7\\
e                                  & 0.24   & 0.02 & 0.00   & 0.0 &
0.32 & 0.04 &0.09       & 0.02 & 0.25    & 0.03 & 0.31   & 0.07\\
$\omega$ ($^{\circ}$)               & 47.   & 7.    &        &     &
160. & 6.  & 133.      & 16.  & 242.     & 6.  & 80.     & 14. \\
T$_{0}$ (periastron) (JD24+)       & 51482. & 10.  &       &     &
538535. & 22. & 51491.     & 17   & 53061. & 8.  & 51849.  & 5. \\
$\gamma$ (\kms)                    & 61.98   & 0.16 & 3.65   & 0.17
&$-$7.1 & 0.4 & $-$14.9   & 0.2  & 67.0 & 0.3 & 33.1  & 0.3 \\
r.m.s.  (\kms)                     & 1.88   &     & 1.47   &     & 2.42 && 2.1 &  &
 1.98 &     & 3.31   &\\  
$a_1 \sin i$ (AU)                  & 0.39   & 0.01 & 0.29 & 0.04    &
2.12 & 0.08  & 0.42      &   0.01   & 0.58  &  0.02  & 0.08  & 0.01 \\
F(M) (M$_{\odot}$)                  & 0.0304   &  0.0033   &   0.0042
&  0.0017   & 0.57 & 0.06 & 0.068    & 0.006     & 0.097 &  0.012   &
0.0041 & 0.0012 \\
N                                  & 161    &     & 82     &     & 63
& & 161        &      & 90   &     & 111     &  \\ 
$\Delta t$ (cycles)                & 6.6     &     & 9.2    &     &3.8
&& 8.7        &      & 5.9  &     & 28.4    &\\
$\Delta m(V)$                      & 0.25   &     & 0.23       &     & 0.81
& & 0.25       &      & 0.19 &     & 0.23   & \\ \hline
\end{tabular}
\end{center}
\end{table*}

\section{Discussion}\label{sect:discussion}

\subsection{Orbits}

The most important conclusion of this extensive monitoring effort is
that we have proven that all the programme stars are binaries.
Not surprisingly, given the high luminosities of the primaries, they are 
single-lined spectroscopic binaries, as in none of the spectra have 
we found evidence of a flux contribution from the companion. 

The orbital elements cover a wide range of periods, from 120 to 1500
days, with a range in mass functions of 0.004 to 0.57 M$_{\odot}$.
The semi-major axes of the orbits of the primaries around the centre of
mass cover the range $a\,\sin i$ =  0.08 to 2.1 A.U.  These are
rather small orbits compared to the dimensions of a single 
AGB star. With their present estimated luminosity, these stars were too 
large while on the AGB to fit in their orbits and they must have been subject 
to severe binary interaction.

The physical interpretation of the orbital elements is hampered by the 
unknown inclination.  Provided the evolved component has a canonical 0.6
M$_{\odot}$ left, the minimal masses of the companions are given in
Table~\ref{tab:mass}.

It is, however, very unlikely we observe the objects nearly edge-on. 
To quantify the impact of the inclination on the observed SED and
observed line-of-sight reddening, we used the detailed radiative transfer 
model of the disc around \nulacht \citep{deroo07} and assume that a 
similar model applies to all programme stars.  The 
inner hole is dust free as we assume that dust grains cannot exist above 
the sublimation temperature. The presence of this inner gap allows the 
existence of a vertical boundary directly irradiated by the central star, 
creating a puffed-up geometry \citep{deroo07}. The scale height of the disc must be 
significant in all objects because of the very significant near infrared 
luminosity in the whole sample \citep{deruyter06}. The temperature, density 
and scale height of the disc are calculated based on the condition of 
hydrostatic equilibrium, and iterated to obtain the convergent structure. 
The constraints of the model parameters of the disc around \nulacht came 
from spectrally resolved interferometric data in both the near-IR and the 
N-band \citep{deroo07}.

\begin{table}
\caption{\label{tab:mass} The evaluation of the mass of the invisible companion
  assuming that the evolved star is of 0.6 M$_{\odot}$. The absence of
  strong circumstellar reddening as well as orbital modulation in the
  photometry shows that the minimal mass of the companion is likely
  obtained with an inclination of at most 65$^{\circ}$. }
\begin{center}
\begin{tabular}{lll} \hline\hline\rule[0mm]{0mm}{3mm}
Star  & \multicolumn{2}{c}{Mass of the companion} \\
      & i = 90$^{\circ}$   &   i = 65$^{\circ}$      \\  \hline
\rule[0mm]{0mm}{3mm}
\nulacht  &  0.29  &  0.33 \\
\hd       &  0.13  &  0.15 \\
\entra    &  1.25  &  1.50 \\ 
\eenvijf  &  0.41  &  0.47 \\
\eennegen &  0.49  &  0.56 \\
\eennegenb & 0.13 & 0.15 \\ \hline
\end{tabular}
\end{center}
\end{table}

\begin{figure}
\begin{center}
\resizebox{\hsize}{!}{\includegraphics{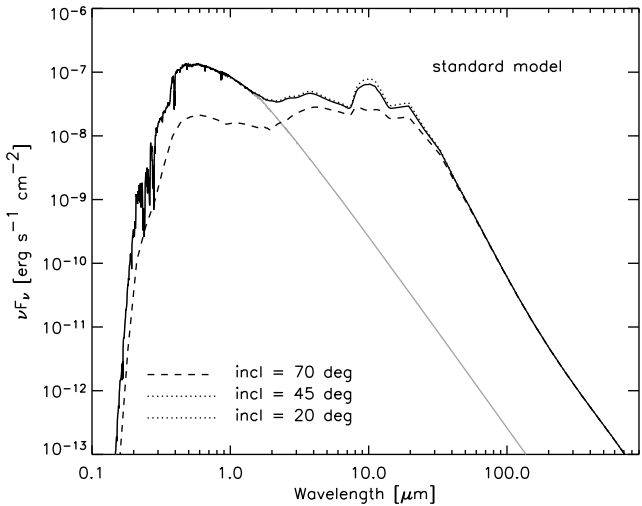}}
\caption{\label{fig:inclination} The impact of the inclination on the 
SED computed by radiative transfer using typical parameters for a passively 
irradiated disc. The lower inclinations coincide on this scale so the
same dotted line is used.}
\end{center}
\end{figure}

In Fig~\ref{fig:inclination} we show the impact on the SED of
different inclinations. The figure illustrates that around an
inclination of 70$^{\circ}$, the line-of-sight towards the central object
will graze the disc. This will strongly affect the circumstellar
reddening. The line-of-sight circumstellar reddening in the 70$^{\circ}$
model is about E(B-V) = 0.8. Edge-on, the discs are strongly opaque
and no direct visible light of the central star will reach us. The
objects with the largest reddening are \nulacht, \eenvijf and
\eennegen with E(B-V) = 1.5 $\pm$ 0.3, 1.5 $\pm$0.3 and 1.2 $\pm$ 0.3
respectively \citep{deruyter06}. The low galactic latitudes and the
strong interstellar DIBs in their spectra \citep[Fig. 3 of][]{maas05}
indicate that the greater part is of interstellar origin,
however. We conclude that in all six objects, the inclination of the
is orbit is very likely to be less than or equal to 65$^{\circ}$ and the corresponding
lower limits of the mass estimates are also given in
Table~\ref{tab:mass}. \entra is the only object for which we have
direct evidence that the companion is too massive to be a white dwarf. Note
that also for the other objects, there is no evidence of a hot white dwarf,
neither in the energetics nor in the lack of spectroscopic evidence of
symbiotic activity. The companion stars are likely to be unevolved
main-sequence stars, which do not contribute significantly to the
energy budget of the systems.

\subsection{Chemical Composition}

A surprising effect of the presence of a disc is that the photospheric
chemical content can be strongly affected by
depletion \citep{vanwinckel92,waters92}. The observed
chemical pattern in the photospheres is the result of gas-dust
separation followed by re-accretion of only the gas, which is poor in
refractory elements. The photospheres are then deficient in
refractories (like Fe and Ca and the s-process elements), while the
non-refractories are not affected. Photospheric depletion is
surprisingly common in evolved objects
\citep[][and references therein]{giridhar05,maas05,maas07}. Recently we
found that depletion is also present in the LMC \citep{reyniers07b}, 
again around objects where the presence of a
gravitationally bound disc is suspected.

Here we strengthen the link between the efficiency of the depletion
process and the binary nature of the object. \nulacht, \eenvijf,
\eennegen and,  to a much lesser extent,  \eennegenb are affected by
photospheric depletion \citep{maas05}. \entra is iron deficient but
the signature of depletion is less clear while \hd has mainly solar
abundances \citep{vanwinckel97}. 
The depletion process will mask third dredge-up
chemical enrichment as the s-process elements are strongly refractory
\citep{lodders03}. In none of the objects is there any evidence for
AGB nucleosynthetic enrichment.

Recent infrared spectra of \entra, \eennegen, \eennegenb were
presented by \citet{gielen08}, for \nulacht by \citet{deroo07}. In all objects the
dustfeatures are dominated by silicates and characterised by a very
strong degree of processing both in grainsize and in
crystallinity. The dust disc creation occurred when the envelope was
O-rich. This is a general characteristic of the whole sample
\citep{vanwinckel03} and the chemical evidence both of the
photospheres and of the circumstellar material show that the AGB evolution of
those binaries was cut short by the phase of strong binary interaction.

\subsection{Binary evolution}

Calculations from \citet{schuerman72} show that, when one takes into
account the radiation pressure of the AGB-star in a close binary
system, the critical Roche potential degenerates into a surface
containing the inner Lagrangian point L$_{1}$ and the outer Lagrangian
point L$_{2}$ for a critical ratio of the radiation pressure force to
the gravitational attraction.  For this critical ratio, material lost
by the AGB star flows through the L$_{1}$ point and can settle itself
through the L$_{2}$ point in a circumbinary 
disc \citep[see also][]{frankowski07}.

Recent population synthesis experiments \citep{axel08} account for
tidally enhanced mass-loss \citep{tout88} during AGB evolution and
balance the eccentricity pumping mechanisms induced by a prescription
of the mass-loss which is orbital phase related, so that at periastron
passage, mass loss is enhanced \citep{soker00b}. The conclusion of
\citet{axel08} was that the eccentricity enhancement process is in
balance with the circularisation for objects which do not enter the
common envelope phase and are expected to leave the AGB in orbits
around 1000 days or longer. The eccentric orbits with shorter periods
(those of \nulacht, \eenvijf, \eennegen and especially \eennegenb) are
still not accounted for. In these systems, additional eccentricity
pumping mechanisms should be at work which may be linked to the
presence of the circumbinary disc \citep{waelkens96,artymowicz91}.
Similar but less pronounced excess eccentricities with respect to
theory are found in related objects \citep[e.g.][]{jorissen09}.
Despite the phase of strong binary interaction, there is no evidence
that the mass transfer towards the undetected companion has been very
efficient: the mass functions implying rather low minimal masses of
the secondary (see Table~\ref{tab:mass}).

The global picture that emerges from our radial velocity monitoring is
that the post-AGB stars evolved in a
system which is too small to accommodate a peak AGB
star. During a badly understood phase of strong interaction, which in view 
of the small orbits was probably a common envelope phase, a
circumbinary dusty disc was formed, but the binary system did not
suffer dramatic decrease in orbital radius.  
What we observe now is an F-G supergiant in a binary 
system, which is surrounded by a circumbinary dusty disc
in a bound orbit. Although not yet proven, the fact that we found a
100 \%  binarity rate for our subsample of objects with low
pulsational amplitude strongly favours the binary nature of all
post-AGB stars with a hot dust component.

The formation, structure and evolution of the disc are
far from being understood, but do appear to be key ingredients
in our understanding of the late evolution of a very significant
binary population.

\subsection{Variability}

These six stars show a wide range of properties, in particular as
regards the amplitude and more especially the regularity of
variation. The first three stars, \nulacht, \hd and \entra, have
periodograms with peaks which stand out strongly from the noise, while
\eenvijf has a much weaker peak and \eennegen and \eennegenb have only
a confused scatter, with minor peaks on a noisy background. \nulacht
is the only one of the six to exhibit well defined, coherent
variations for as long as one or two years. \hd and \entra have better
defined peaks in the periodogram than \nulacht, with a single period
in the case of \entra, but their amplitudes and, in the case of \hd,
the lengths of individual cycles change continuously. The variations
of \eenvijf and \eennegen exhibit cycles of very variable amplitude
and length, with evidence of changes of direction of pulsation within
a given cycle for \eennegen, while the data for \eennegenb are
inadequate to elucidate the nature of the variations, which are
probably even more irregular and of relatively short period.

These stars are of earlier spectral type and so are hotter than
classical RV Tauri stars \citep{lloydevans99} and appear therefore to
lie to the blue of the Cepheid instability strip. This raises the
question of the nature of the pulsation.  
The ratio of the two principal periods is 0.96 for \nulacht and
0.91 for \hd, whereas the ratio of the first overtone to
the fundamental, P1/P0, is always close to 0.705 in classical Cepheids
\citep{stobie77}. Two high overtones might have the period ratios which
we observe, but it seems implausible that these would be dominant, and
the periods of these two stars are quite long, implying very long
fundamental periods and a correspondingly high luminosity. This
suggests that the pulsations of these stars are non-radial. EN TrA,
which has a single, shorter, period and a larger amplitude, is
probably a radial pulsator, however.
 The apparently abrupt turn round in the direction of change of V
magnitude, leading to a second maximum and a consequent lengthening of
individual cycles in \eennegen, recalls the suggestion that shockwaves
may be involved in the complex motions of the outer atmosphere of the
post-AGB star HD 56126 \citep{barthes00}.

The four stars of largest amplitude showed indications of the
classical RV Tauri feature of alternating deep and shallow minima, but
with equally large alternation of the magnitude at successive maxima.

Multicolour observations may be used to compare light and colour
curves, which follow a similar phase relationship in classical
Cepheids but which differ in Type II Cepheids, for which an
atmospheric shock wave leads to a bluer colour on the rise to maximum
light, with wide loops in two-colour diagrams, whereas classical
Cepheids have very narrow loops. The amount by which the B curve leads
the V curve, in terms of relative phase in the single period, is:
\nulacht, 0.00:, \hd, 0.04, \entra, 0.17. The scattered data and small
amplitudes preclude any finding for the remaining stars. It is
unfortunate that the SAAO observations of \nulacht were made when the
amplitude was relatively small. It is only for \entra, the star of
largest amplitude, that the phase shift typical of Type II Cepheids is
clearly seen.

\section{Conclusions}

The summary of the main conclusions of this contribution is that:

\begin{itemize}
\item Our radial velocity monitoring show that post-AGB stars which combine a
  hot dust component (and hence have a disc) and a low photospheric pulsational 
  amplitude  \citep{deruyter06} are binaries. Even in these objects, the
  amplitude of the radial velocity variations induced by the pulsation
  is a significant fraction of the orbital velocity. It is an
  observational challenge to prove the binary nature of all post-AGB
  objects with a hot dust component, as in many systems large
  amplitude pulsations are present. Nevertheless the fact that 100 \%
  of our low amplitude pulsator programme stars are binaries is a
  significant indication that binarity is a prerequisite for disc formation.

\item We confirm and strengthen the conclusion of \cite{kiss07}
  that complex multi-periodic but low-amplitude pulsational behaviour is
  present at the blue side of the Type II Cepheid instability strip.

\item The evolution of the binaries in the resulting period-eccentricity 
  parameter space is not understood using standard binary
  evolution channels. Recent theoretical efforts include enhanced
  mass-loss and an efficient eccentricity pumping mechanism induced
  by modulation of the mass-loss in the orbital period. Nevertheless
  the high eccentricity of  \nulacht,  \eennegen and especially
  \eennegenb (with a period of 120 days and a significant non-zero
  eccentricity) are detected in systems that are very likely
  post-common envelope systems and in these systems, the high eccentricity
  remains very poorly understood. The on average small mass functions
  imply that mass transfer to the companion has been limited.

\item In all systems found so far, the dust in the disc is O-rich.
  The AGB evolution was likely cut short by the binary interaction event
  and with the orbital elements found in this work, the systems should be seen as 
  post-interacting binaries rather than canonical post-AGB stars.

\item  The wide range of orbits and mass function detected here show
  that the creation of a circumbinary gravitationally bound disc is
  the final evolutionary product of a wide
  range of binaries.

\end{itemize}
\begin{acknowledgements}
The authors want to acknowledge the Geneva Observatory and its staff
for the generous time allocation on the Swiss Euler telescope.  The
IvS acknowledges support from the Fund for Scientific Research of
Flanders (FWO) under the grants G.0178.02., G.0703.08, G.0332.06 and
G.0470.07.  This research was made possible thanks to support from
the Research Council of K.U.Leuven under grant GOA/2008/04.  We
would like to express our appreciation to Grzegorz Pojmanski for his
efforts to provide and maintain the ASAS data in a form accessible
to the astronomical community.

\end{acknowledgements}
\bibliographystyle{aa}

\end{document}